\documentstyle[aps,prl,preprint,floats,epsfig]{revtex} 
   
\newcommand{\ra}{\rightarrow}
\newcommand{\lam}{\Lambda^0}

\newcommand{\sig}{\Sigma^+}
\newcommand{\si}{\Sigma^0}

\begin{document}

\preprint{\tighten\vbox{\hbox{\hfil BELLE-CONF-0130}}}

\title{
\quad\\[3cm] \Large  Observation of $\Lambda_c^+\ra \lam K^+$, $\Lambda_c^+\ra \si K^+$ and \\[-10pt]  
$\Lambda_c^+\ra \Sigma^+ K^+\pi^-$ decays
}

\author{The Belle Collaboration}

\maketitle

\tighten


\begin{abstract}

We present measurements of the Cabibbo-suppressed decays  
$\Lambda_c^+\ra \lam K^+$ and $\Lambda_c^+\ra \si K^+$
(both first observations), $\Lambda_c^+\ra \Sigma^+ K^+\pi^-$ 
(seen with large statistics for the first time), 
$\Lambda_c^+ \ra p K^+ K^-$ and $\Lambda_c^+\ra p \phi$
(measured with improved accuracy).
Improved branching ratio measurements for the decays
$\Lambda_c^+ \ra \sig K^+ K^-$ and $\Lambda_c^+\ra \sig \phi$,
which are attributed to W-exchange diagrams, are shown. 
We also present the first evidence for $\Lambda_c^+ \ra \Xi(1690) K^+$ 
and set an upper limit on non-resonant $\Lambda_c^+ \ra \sig K^+ K^-$ decay.
This analysis was performed using 23.6~fb$^{-1}$ of data collected 
by the Belle detector at the $e^+ e^-$ asymmetric collider KEKB. 

\end{abstract}

%
		

{\renewcommand{\thefootnote}{\fnsymbol{footnote}}

\setcounter{footnote}{0}

\newpage


\normalsize


\begin{center}
  K.~Abe$^{9}$,               
  K.~Abe$^{37}$,              
  R.~Abe$^{27}$,              
  I.~Adachi$^{9}$,            
  Byoung~Sup~Ahn$^{15}$,      
  H.~Aihara$^{39}$,           
  M.~Akatsu$^{20}$,           
  K.~Asai$^{21}$,             
  M.~Asai$^{10}$,             
  Y.~Asano$^{44}$
% ...
\end{center}

\small
\begin{center}
$^{1}${Aomori University, Aomori}\\
% ...
$^{9}${High Energy Accelerator Research Organization (KEK), Tsukuba}\\
$^{10}${Hiroshima Institute of Technology, Hiroshima}\\
% ...
$^{13}${Institute for Theoretical and Experimental Physics, Moscow}\\
$^{14}${Kanagawa University, Yokohama}\\
$^{15}${Korea University, Seoul}\\
% ...
$^{20}${Nagoya University, Nagoya}\\
$^{21}${Nara Women's University, Nara}\\
% ...
$^{27}${Niigata University, Niigata}\\
% ...
$^{30}${Panjab University, Chandigarh}\\
% ...
$^{37}${Tohoku Gakuin University, Tagajo}\\
% ...
$^{39}${University of Tokyo, Tokyo}\\
% ...
$^{44}${University of Tsukuba, Tsukuba}\\
% ...
\end{center}

\normalsize

\newpage


\section{Introduction}

Decays of charmed baryons, unlike charmed mesons, 
are not colour or helicity suppressed, allowing us to investigate the
contribution of W-exchange diagrams. 
There are also possible interference effects due to the presence of identical
quarks. This makes the study of these decays a useful tool to test 
theoretical models that predict exclusive decay rates~\cite{theory}.

During the past several years there has been significant progress in the
experimental study of hadronic decays of charmed baryons.
New results on masses, widths, lifetimes and asymmetry 
decay parameters have been published by various experiments~\cite{PDG}.
However the accuracy of branching ratio measurements does not exceed 30\% for
many Cabibbo-favoured modes: for Cabibbo-suppressed and W-exchange dominated
decays, the experimental accuracy is even worse.
As a result, we are not yet able to conclusively distinguish between the decay
rate predictions made by different theoretical models.

In this paper we present a study of $\Lambda_c^+$ baryons produced in the
$e^+ e^- \ra q {\bar q}$ continuum at Belle, relying on the excellent particle
identification system of the detector to measure decays with 
kaons in the final state.
We report the first observation of the Cabibbo-suppressed decays
$\Lambda_c^+ \ra \lam K^+$ and $\Lambda_c^+ \ra \si K^+$, and the
first observation of $\Lambda_c^+ \ra \sig K^+ \pi^-$ with large statistics.
(Here and throughout this paper, the inclusion of charge-conjugate states is
implied.)
We present improved measurements of the Cabibbo-suppressed decays
$\Lambda_c^+ \ra p K^+ K^-$ and $\Lambda_c^+\ra p \phi$,
and the W-exchange decays  $\Lambda_c^+ \ra \sig K^+ K^-$ and  
$\Lambda_c^+\ra \sig \phi$;
we also report the first evidence for $\Lambda_c^+ \ra \Xi(1690) K^+$,
and set an upper limit on non-resonant $\Lambda_c^+ \ra \sig K^+ K^-$ decay.
All branching ratio measurements and limits are preliminary.


\section{Data and Selection Criteria}
\label{section-selection}

The data used for this analysis were taken on the $\Upsilon(4S)$ 
resonance and in the nearby continuum
using the Belle detector at the $e^+ e^-$ asymmetric collider KEKB. 
The integrated luminosity of the data sample is equal to 23.6~fb$^{-1}$. 

Belle is a general purpose detector based on a 1.5 T superconducting solenoid;
a detailed description can be found elsewhere~\cite{BELLE_DETECTOR}.
Tracking is performed by a silicon vertex detector (SVD) composed of three
concentric layers of double sided silicon strip detectors, and a 50 layer
drift chamber.
Particle identification for charged hadrons, important for the measurement of
final states with kaons and/or protons,  is based on the combination of  
energy loss measurements $(dE/dx)$ in the drift chamber,
time of flight measurements and aerogel {\v C}erenkov counter information. 
For each charged track, measurements from these three subdetectors are 
combined to form $K/\pi$ and $p/K$ likelihood ratios in the range from 0 to 1,\\[-3pt]
$${\rm P}(K/\pi) = {\cal L}(K)/({\cal L}(K) + {\cal L}(\pi)),~~ 
{\rm P}(p/K) = {\cal L}(p)/({\cal L}(p) + {\cal L}(K)),$$
where ${\cal L}(p)$, ${\cal L}(K)$ and ${\cal L}(\pi)$ are the likelihood values
assigned to each identification hypothesis for a given track.

For the analyses presented here, we require P($K/\pi) < 0.9$ for pions, 
P($K/\pi) > 0.6$ for kaons, and P($p/K) > 0.9$ for protons unless otherwise
stated.
Candidate $\pi^0$'s are reconstructed from pairs of photons detected in the 
CsI calorimeter, with a minimum energy of 50~MeV per photon.
Other particles are identified as follows:
\begin{itemize}
  \item	$\lam$ are reconstructed in the decay mode $\lam \rightarrow p \pi^-$,
	fitting the $p$ and $\pi$ tracks to a common vertex and  
	requiring an invariant mass in a $\pm 3$MeV$/c^2$ ($\approx 3\sigma$)
	interval around the nominal value.
	The likelihood ratio cut on the proton is relaxed to P$(p/K) > 0.4$.
	We then make the following cuts on the $\lam$ decay vertex: 
	\begin{itemize}
	  \item	the difference Z$_{\rm dist}$ in z-coordinate between
		the proton and pion at the decay vertex must satisfy
		Z$_{\rm dist}<1$~cm;
	  \item	the distance between the vertex position and interaction point
		(IP) in the $r-\phi$ plane must be $> 1$~mm;
	  \item	the angle $\alpha$, between the $\lam$ momentum vector and the
		vector pointing from the IP to the decay vertex,
		must satisfy $\cos\alpha > 0.995$.
	\end{itemize}
	Finally, to reduce the combinatorial background, we require the $\lam$ 
	momentum to be greater than $1.5\, \mathrm{GeV}/c$.
  \item	$\sig$ are reconstructed in the decay mode $\sig \rightarrow p \pi^0$,
	requiring an invariant mass within $\pm 10\, \mathrm{MeV}/c^2$
	($\approx 2.0 \sigma$) of the nominal value.
	Since the $\sig$ mass resolution is dominated by the uncertainty in the
	$\pi^0$ momentum, rather than the choice of $\sig$ decay vertex, we
	use the IP to estimate the vertex for the $\pi^0$ fit; this
	approximation has been checked in the Monte Carlo.
	The displaced $\sig$ vertex is taken into account using
	the impact parameter of the decay proton in the $r-\phi$ plane
	with respect to the IP, $d_{r\phi}$.
	We require the proton to have at least one hit in the SVD,
	to improve the impact parameter resolution, and then make a cut
	$d_{r\phi} > 500\,\mu\mathrm{m}$.
  \item	$\si \rightarrow \lam \gamma$ decays are formed using identified $\lam$
	and photons with calorimeter cluster energies $E_\gamma > 0.1$ GeV;
	we accept candidates with invariant masses within $\pm 6$~MeV$/c^2$
	($\approx 1.5\sigma$) of the nominal value.
	To suppress the background from slow photons, the cosine 
	of the angle between the $\gamma$ in the $\si$ rest frame and the $\si$ 
	boost direction is required to be greater than $-0.8$. 
\end{itemize}

To suppress combinatorial and $B\overline{B}$ backgrounds,
we require $\Lambda_c^+$ candidates to have scaled momentum
$x_p=p^\ast(\Lambda_c^+)/\sqrt{E_{CMS}^2/4-M^2} >0.5$, where 
$p^\ast(\Lambda_c^+)$ and $E_{CMS}$ are the candidate momentum and total
$e^+e^-$ beam energy in the center of mass frame, respectively, and $M$ the
reconstructed mass of the $\Lambda_c^+$.
In modes where there are two or more charged tracks at the $\Lambda_c^+$
vertex, we perform a vertex fit and require $\chi^2/\mathrm{n.d.f.} < 9$.


\section{First observation of the decays
	$\Lambda_{\lowercase {c}}^+\ra \lam K^+$ and
	$\Lambda_{\lowercase {c}}^+\ra \Sigma^0 K^+$}

The Cabibbo-suppressed decay $\Lambda_{\lowercase {c}}^+\ra \lam K^+$ has not
previously been observed. Reconstructing $\lam K^+$ combinations as described
in section~\ref{section-selection}, we see a clear signal at the $\Lambda_c^+$
mass, as shown in Fig.~\ref{fig_lamc_lam0k_1}(a).

\begin{figure}[p]
\centering
\begin{picture}(450,470)
\put(-15,480){\large $\frac{N}{5.0~{\rm MeV}/c^2}$} 
\put(-15,220){\large $\frac{N}{5.0~{\rm MeV}/c^2}$} 
\put(340,460){\large (a)} 
\put(340,200){\large (b)} 
\put(145,15){$M(\lam K^+)-M(\lam)+1.1157,~{\rm GeV}/c^2$} 
\put(55,290){\includegraphics[width=0.75\textwidth]{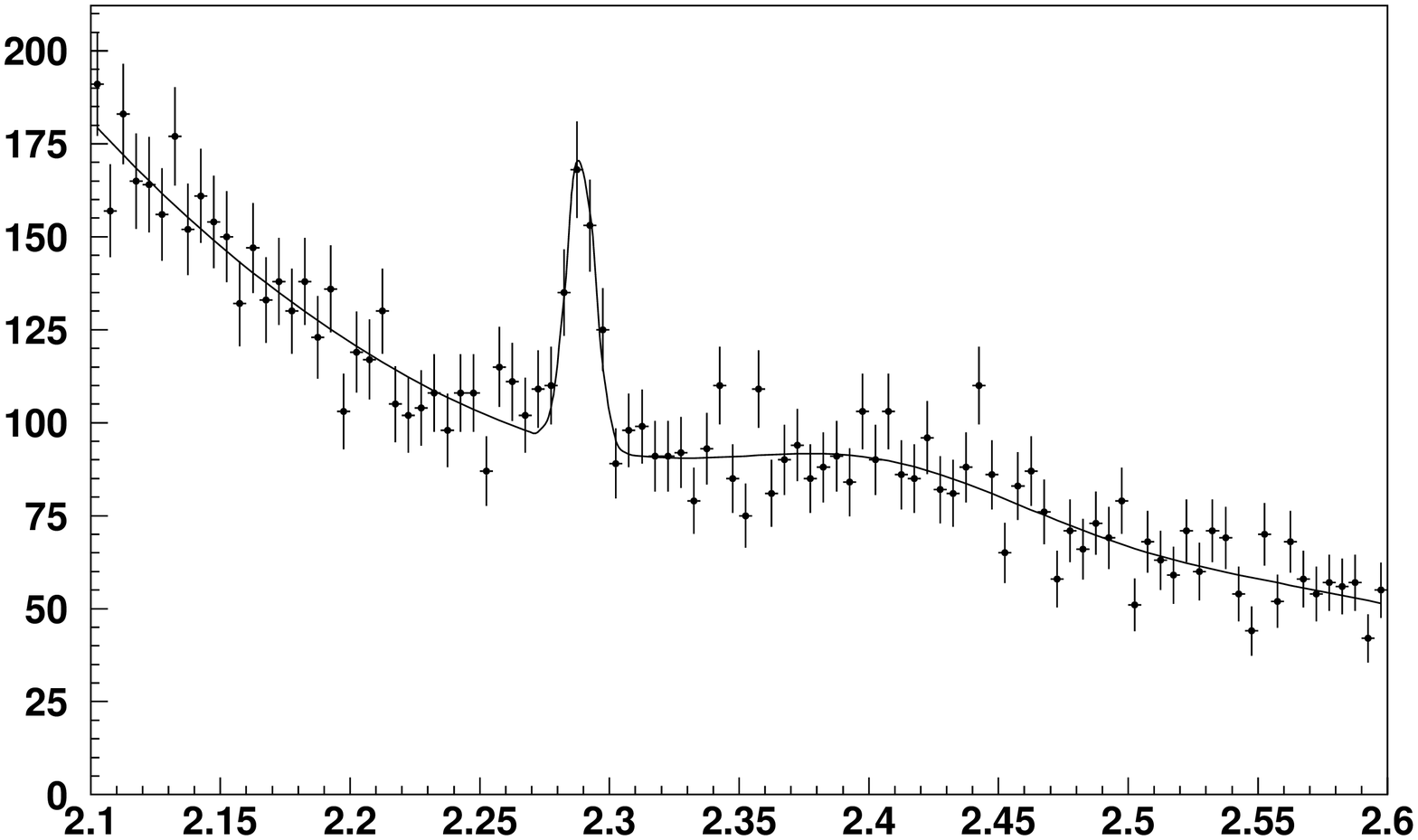}}
\put(55,30){\includegraphics[width=0.75\textwidth]{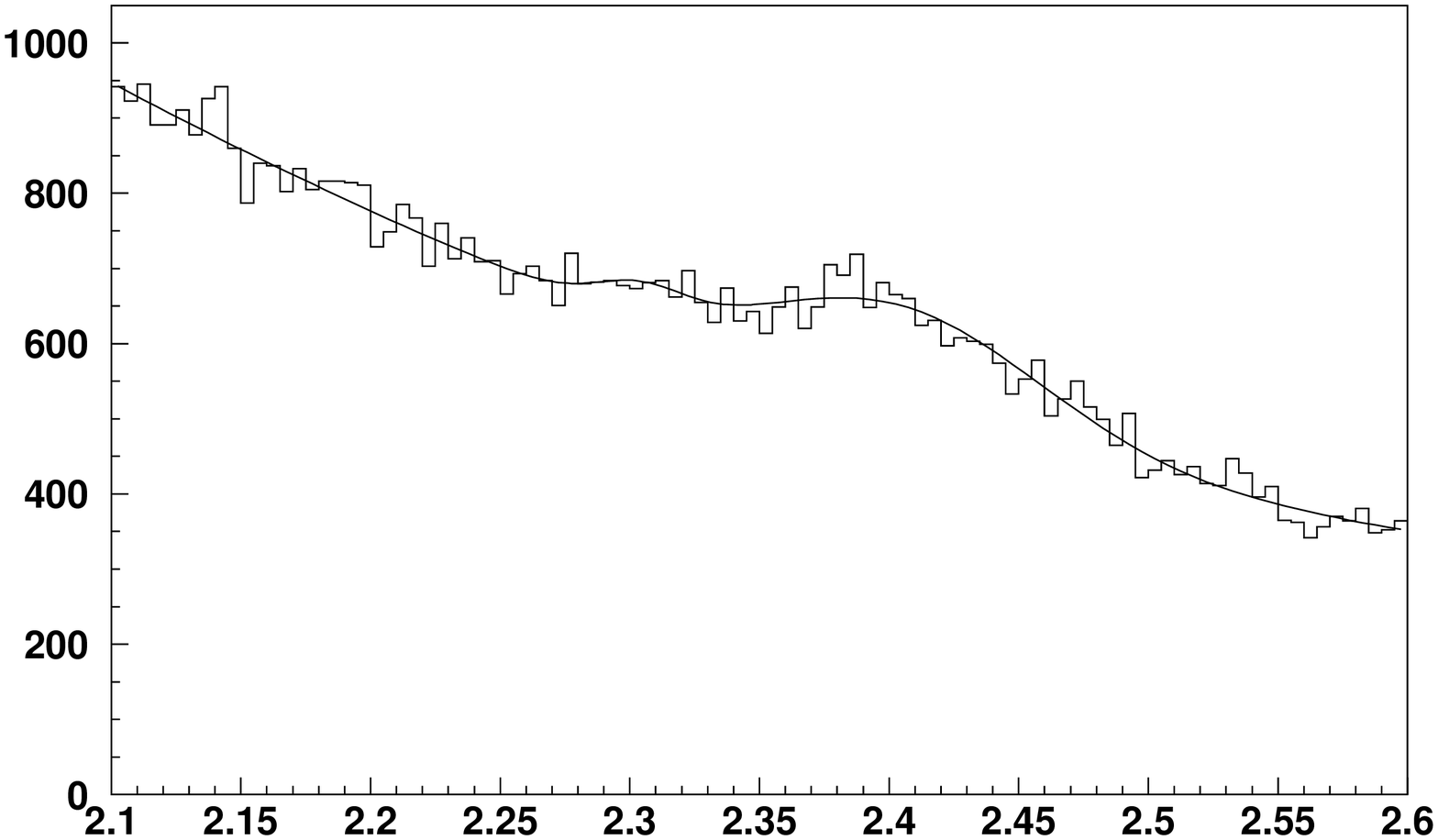}}
\end{picture}
\caption{$\Lambda_{\lowercase {c}}^+\ra \lam K^+$: invariant mass spectra of
	the selected $\lam K^+$ combinations. 
	Plot (a) is made with the requirement P(K/$\pi)>0.6$ on the kaon
	candidate. 
	Plot (b) is made with the requirement P(K/$\pi)<0.1$ 
	to get the form of the contribution from misidentified pions.
	The other selection requirements and the fitting procedure are
	described in the text.
	The structure near $2.4\, \mathrm{GeV}/c^2$ is due to misidentified 
	$\Lambda_c^+\ra \lam \pi^+$ and $\Lambda_c^+\ra \Sigma^0 \pi^+$ decays.}
\label{fig_lamc_lam0k_1}
\end{figure}

To study backgrounds due to Cabibbo-allowed decays, we reconstruct a second
sample with a tight pion identification cut P$(K/\pi)<0.1$ applied to the
``kaon'', Fig.~\ref{fig_lamc_lam0k_1}(b); the kaon mass hypothesis is still
used in this case.
We see a broad structure centered around $2400\, \mathrm{MeV}/c^2$:
using the MC simulation we find that this is produced by Cabibbo-allowed
$\Lambda_c^+\ra \lam \pi^+$ and $\Lambda_c^+\ra \Sigma^0 \pi^+$ decays. 
The background due to $\Lambda_c^+ \ra \Sigma^0 \pi^+$ is dangerous, since the
mass shifts due to the undetected $\gamma$ (from $\Sigma^0\ra \lam \gamma$)
and $\pi \rightarrow K$ misidentification partially cancel, leaving a peak close
to the $\Lambda_c^+$ mass.  
Feed-downs from other Cabibbo-allowed $\Lambda_c$ decays,
$\Lambda_c^+ \ra \lam \pi^+\pi^0$,
$\Lambda_c^+ \ra \lam \pi^+\pi^+\pi^-$ and
$\Lambda_c^+ \ra \Sigma^0 \pi^+ \pi^0$,
produce broad reflections due to the missing particle(s), and no prominent
structures are seen. We therefore fit the distribution
using two gaussians with floating central values and widths 
(to model the $\Lambda_c^+\ra \lam \pi^+$ and $\Lambda_c^+\ra \Sigma^0 \pi^+$
contributions), and a second order polynomial (to model the wide reflections
and the remaining background); the result of the fit is superimposed on
Fig.~\ref{fig_lamc_lam0k_1}(b).
All parameters except for the overall normalization are then fixed,
and we use this function to model the $\lam \pi^+$ background in the main
sample (Fig.~\ref{fig_lamc_lam0k_1}(a)).

The remaining combinatorial background in Fig.~\ref{fig_lamc_lam0k_1}(a) is 
represented using a second order polynomial, and the 
$\Lambda_c^+\ra \lam K^+$ signal is described by a gaussian 
with width $\sigma=5.4\, \mathrm{MeV}/c^2$ (fixed from MC);
the result of the fit is shown by the superimposed curve.
We find a yield of $214 \pm 30$ $\Lambda_c^+\ra \lam K^+$ decays,
the first observation of this decay mode. 

\begin{figure}[b!]
\centering
\begin{picture}(550,280)
\put(-25,230){\large $\frac{N}{5.0~{\rm MeV}/c^2}$} 
\put(150,20){$M(\lam \pi^+)-M(\lam)+1.1157,~{\rm GeV}/c^2$} 
\put(45,30){\includegraphics[width=0.85\textwidth]{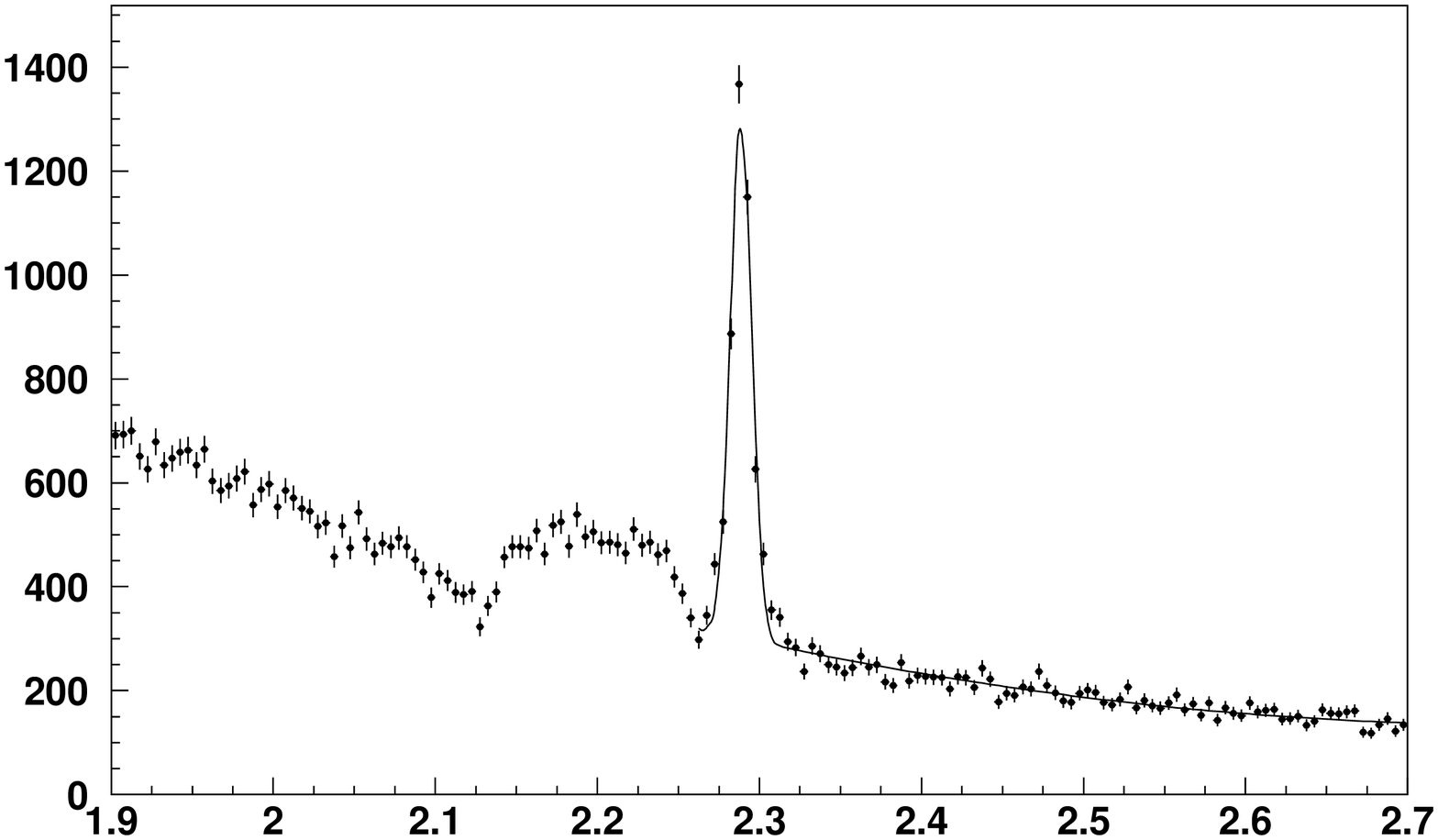}}
\end{picture}
\caption{The invariant mass spectrum for the normalization mode
	$\Lambda_c^+\ra \lam\pi^+$. The selection requirements and fit 
	are described in the text.
	The broad feature below the $\Lambda_c^+$ mass is due to
	$\Lambda_c^+\ra \Sigma^0 \pi^+$ decays.}
\label{fig_lamc_lam0k_11}
\end{figure}

For normalization, we use the decay $\Lambda_c^+\ra \lam\pi^+$: the mass
distribution of the candidates is shown in Fig.~\ref{fig_lamc_lam0k_11}.
A fit with a gaussian with a floating width for the signal and 
a second order polynomial for the background
(restricted to the region above the $\Lambda_c^+ \ra \si \pi^+$ reflection)
yields $3270 \pm 95$ events; 
the fitted signal width $\sigma=6.6\pm 0.2$~MeV$/c^2$ is consistent with the MC
prediction of 6.8~MeV$/c^2$. 
The relative reconstruction efficiency is found to be 
$\epsilon(\Lambda_c^+\ra \lam K^+)/\epsilon(\Lambda_c^+\ra \lam\pi^+) 
 = 0.18/0.23 = 0.77$ in the Monte Carlo: using this value, we extract the 
branching ratio
\[
   {\cal B}(\Lambda_c^+\ra \lam K^+)
  /{\cal B}(\Lambda_c^+\ra \lam\pi^+)=0.085\pm 0.012 \pm 0.015.
\] 
The first error is statistical, and the second is systematic, 
combining the effects of uncertainties in $K/\pi$ identification 
efficiencies and the result of varying the ranges and parameters of the fits.

\begin{figure}[b!]
\centering
\begin{picture}(550,240)
\put(30,190){\large $\frac{N}{5.0~{\rm MeV}/c^2}$} 
\put(165,20){$M(\Sigma^0 K^+)-M(\Sigma^0)+1.192,~{\rm GeV}/c^2$} 
\put(90,25){\includegraphics[width=0.7\textwidth]{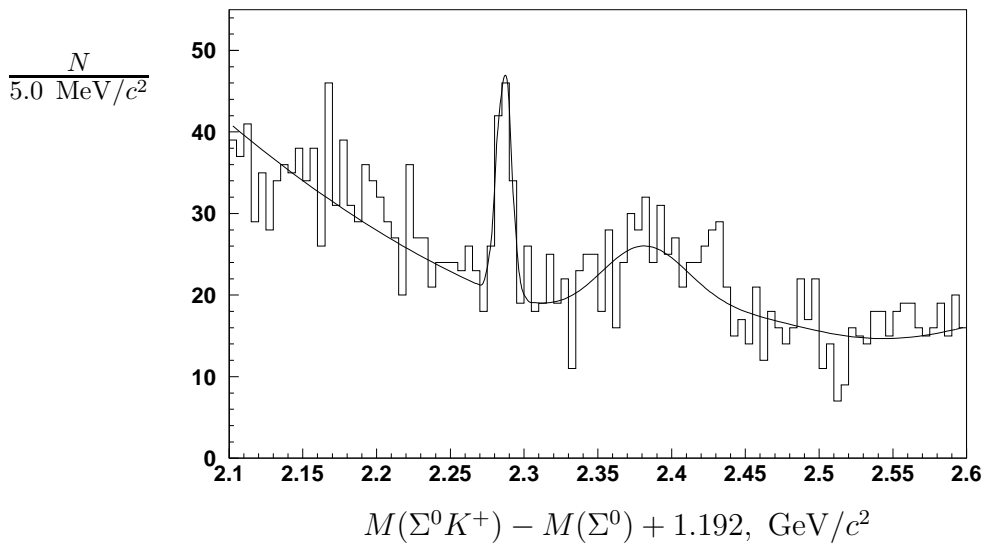}}
\end{picture}
\caption{$\Lambda_c^+ \ra \Sigma^0 K^+$: invariant mass spectrum of the
	selected $\Sigma^0 K^+$ combinations. 
	The selection requirements and fit are described in the text.}
\label{fig_lamc_sig0k_1}
\end{figure}

The Cabibbo-suppressed decay $\Lambda_c^+\ra \Sigma^0 K^+$ is reconstructed in
a similar way, with the scaled momentum cut tightened to $x_p>0.6$ to suppress
the large background due to soft photons. The invariant mass distribution of
the selected $\si K^+$ candidates is shown in Fig.~3:
a peak is seen at the $\Lambda_c^+$ mass, and a reflection due to misidentified
two-body Cabibbo-allowed $\Lambda_c^+$ decays is seen at higher masses. 
The superimposed curve shows the result of a fit constructed using the 
method described for $\lam K^+$, with the exception that the width of the
signal gaussian is fixed from the Monte Carlo to 
$\sigma=5.0$~MeV$/c^2$ in this case.
We find $70\pm 17$ $\Lambda_c^+\ra \si K^+$ events, the first observation of
this decay mode. 

\begin{figure}[h]
\centering
\begin{picture}(550,200)
\put(30,190){\large $\frac{N}{5.0~{\rm MeV}/c^2}$} 
\put(165,20){$M(\Sigma^0 \pi^+)-M(\Sigma^0)+1.192,~{\rm GeV}/c^2$} 
\put(90,25){\includegraphics[width=0.7\textwidth]{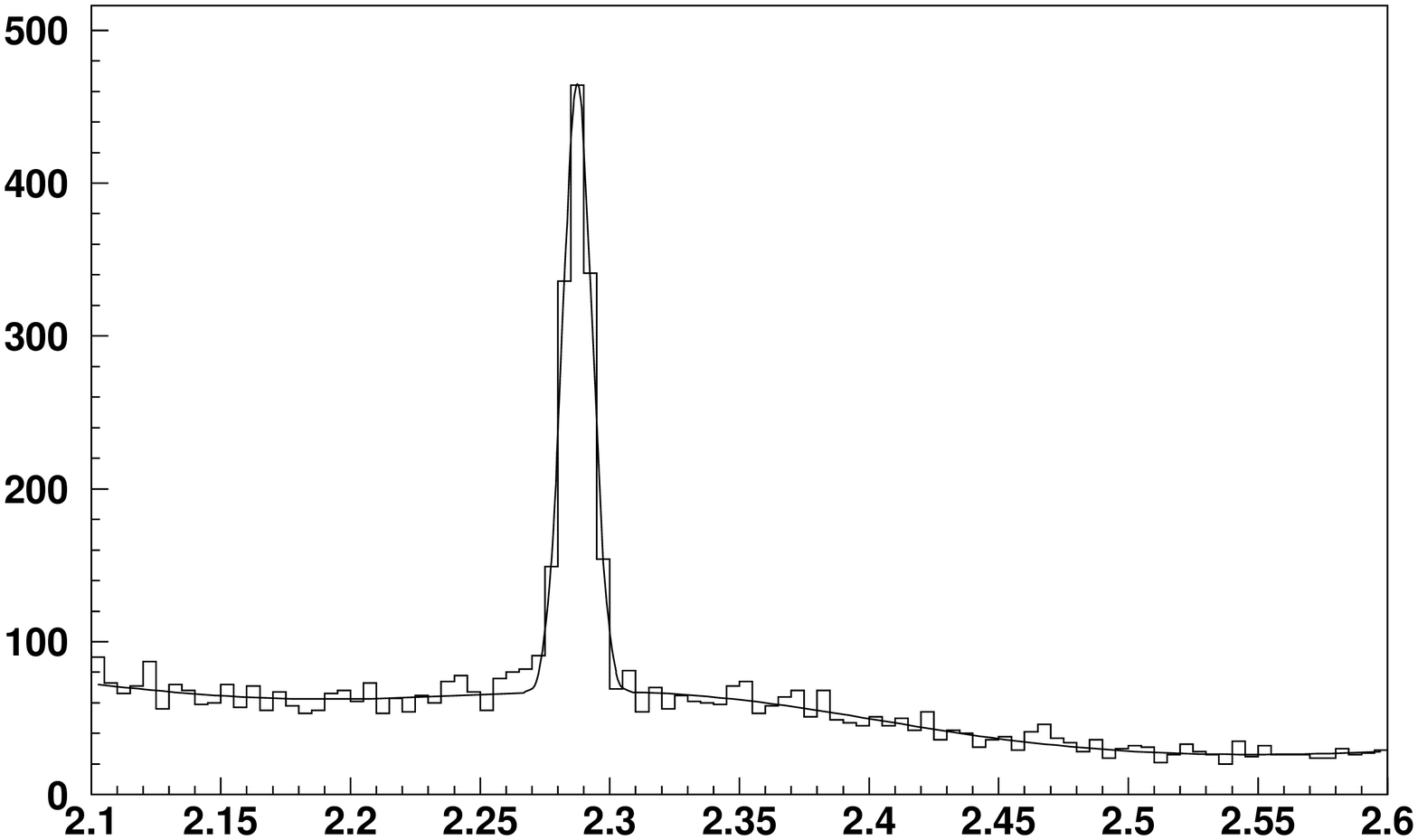}}
\end{picture}
\caption{The invariant mass spectrum for the normalization mode
	$\Lambda_c^+ \ra \Sigma^0 \pi^+$. The selection requirements and fit 
	are described in the text.
	The broad enhancement around the $\Lambda_c^+$ mass region is due to
	$\Lambda_c^+\ra \Lambda^0 \pi^+$ decays.}
\label{fig_lamc_sig0k_11}
\end{figure}

For normalization, we use the decay $\Lambda_c^+\ra \si \pi^+$, 
shown in Fig.~4. We fit the distribution with a gaussian for the signal,
a second gaussian to describe the broad bump due to $\Lambda_c^+\ra \lam \pi^+$
(with the addition of a random $\gamma$), and a second order polynomial for the
remaining background. The mean of the signal gaussian is fixed at the 
$\Lambda_c^+$ mass, and all other parameters are allowed to float:
the fitted width of the signal $\sigma=5.7\pm 0.2$~MeV$/c^2$ 
is consistent with the MC prediction $\sigma=6.1$~MeV$/c^2$.
The fit gives $1132 \pm 39$ $\Lambda_c^+\ra \si \pi^+$ decays. 
The relative reconstruction efficiency is found to be  
$\epsilon(\Lambda_c^+\ra \si K^+)/\epsilon(\Lambda_c^+\ra \si\pi^+) 
 = 0.096/0.114 = 0.84$ in the Monte Carlo: we then extract the branching ratio
\[
   {\cal B}(\Lambda_c^+\ra \si K^+)
  /{\cal B}(\Lambda_c^+\ra \si\pi^+)=0.073\pm 0.018 \pm 0.016.
\] 
The first error is statistical, and the second is systematic, combining
uncertainties from particle identification efficiencies and the fitting
procedure.


\section{Observation of the $\Lambda_{\lowercase {c}}^+ \ra \sig K^+ \pi^-$ decay}

\begin{figure}[b!]
\centering
\begin{picture}(550,220)
\put(30,190){\large $\frac{N}{3.0~{\rm MeV}/c^2}$} 
\put(155,20){$M(\sig K^+ \pi^-)-M(\sig)+1.189,~{\rm GeV}/c^2$} 
\put(90,25){\includegraphics[width=0.7\textwidth]{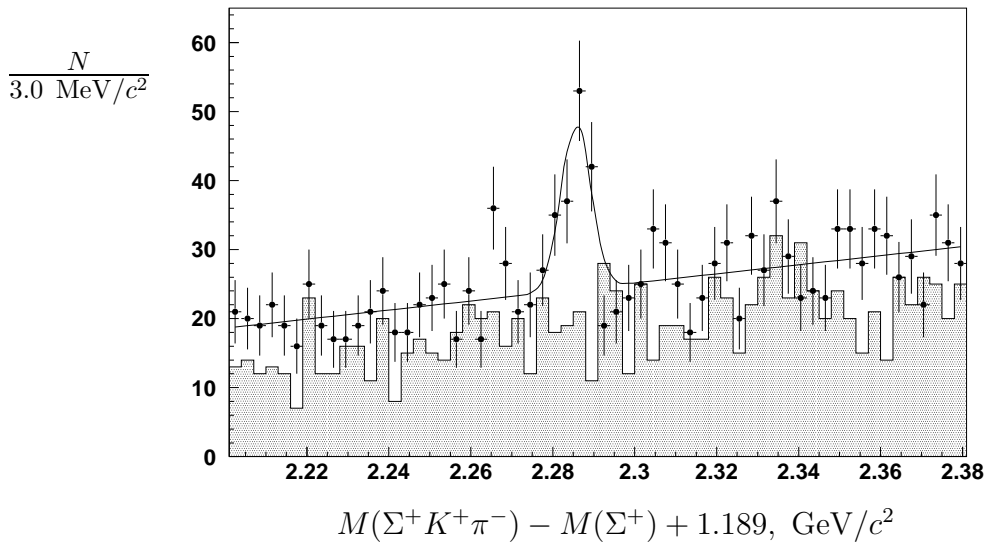}}
\end{picture}
\caption{$\Lambda_c^+ \ra \sig K^+ \pi^ -$: invariant mass spectrum of the
	selected $\sig K^+ \pi^-$  combinations. The shaded histogram
	shows the equivalent spectrum for the $\sig$ sidebands.}
\label{fig_lamc_sigkpi}
\end{figure}

The first evidence for the Cabibbo-suppressed decay $\Lambda_c^+ \ra \sig K^+ \pi^-$
was published by the NA32 collaboration in 1992~\cite{NA32_lamc_sigkpi}:
they found 2 events in the signal region. 
Reconstructing $\sig K^+ \pi^-$ combinations with the cuts of
section~\ref{section-selection} tightened to require P($K/\pi$)$> 0.9$ for the
kaon, and $x_p > 0.6$, we see a clear signal peak at the $\Lambda_c^+$ mass,
as shown in Fig.~\ref{fig_lamc_sigkpi}.
Tighter cuts are used to suppress the large combinatorial background.
We also form $\sig K^+ \pi^-$ combinations using ``$\sig$'' candidates
from mass sidebands ($> 2\sigma$ away from the nominal $\sig$ mass), shown 
with the shaded histogram: no enhancement is seen near the $\Lambda_c^+$ mass.

The mass distribution is fitted with a gaussian 
for the signal (with width fixed to $3.6$~MeV$/c^2$ from the MC) and a first
order polynomial for the background:
we find $72 \pm 16$ $\Lambda_c^+ \ra \sig K^+ \pi^-$ events. 
For normalization we reconstructed $\Lambda_c^+ \ra \sig \pi^+ \pi^-$ decays
with the same cuts, finding $1432\pm 78$ events. 
The relative efficiency of the $\Lambda_c^+ \ra \sig K^+ \pi^-$ channel
reconstruction with respect to $\Lambda_c^+ \ra \sig \pi^+ \pi^-$ 
was found to be $0.046/0.055 = 0.85$ in the Monte Carlo: using this value,
we extract the branching ratio
\[ 
   {\cal B}(\Lambda_c^+ \ra \sig K^+ \pi^-)
  /{\cal B}(\Lambda_c^+ \ra \sig \pi^+ \pi^-) = 0.059 \pm 0.014 \pm 0.006;
\]
the systematic error (quoted second) is dominated by the uncertainty in the
relative $K/\pi$ identification efficiency.


\section{Measurement of the $\Lambda_{\lowercase {c}}^+ \ra \sig K^+ K^-$ and 
$\Lambda_{\lowercase {c}}^+ \ra \sig \phi$ decays}

\begin{figure}[p]
\centering
\begin{picture}(550,220)
\put(30,190){\large $\frac{N}{2.0~{\rm MeV}/c^2}$} 
\put(155,20){$M(\sig K^+ K^-)-M(\sig)+1.189,~{\rm GeV}/c^2$} 
\put(90,25){\includegraphics[width=0.7\textwidth]{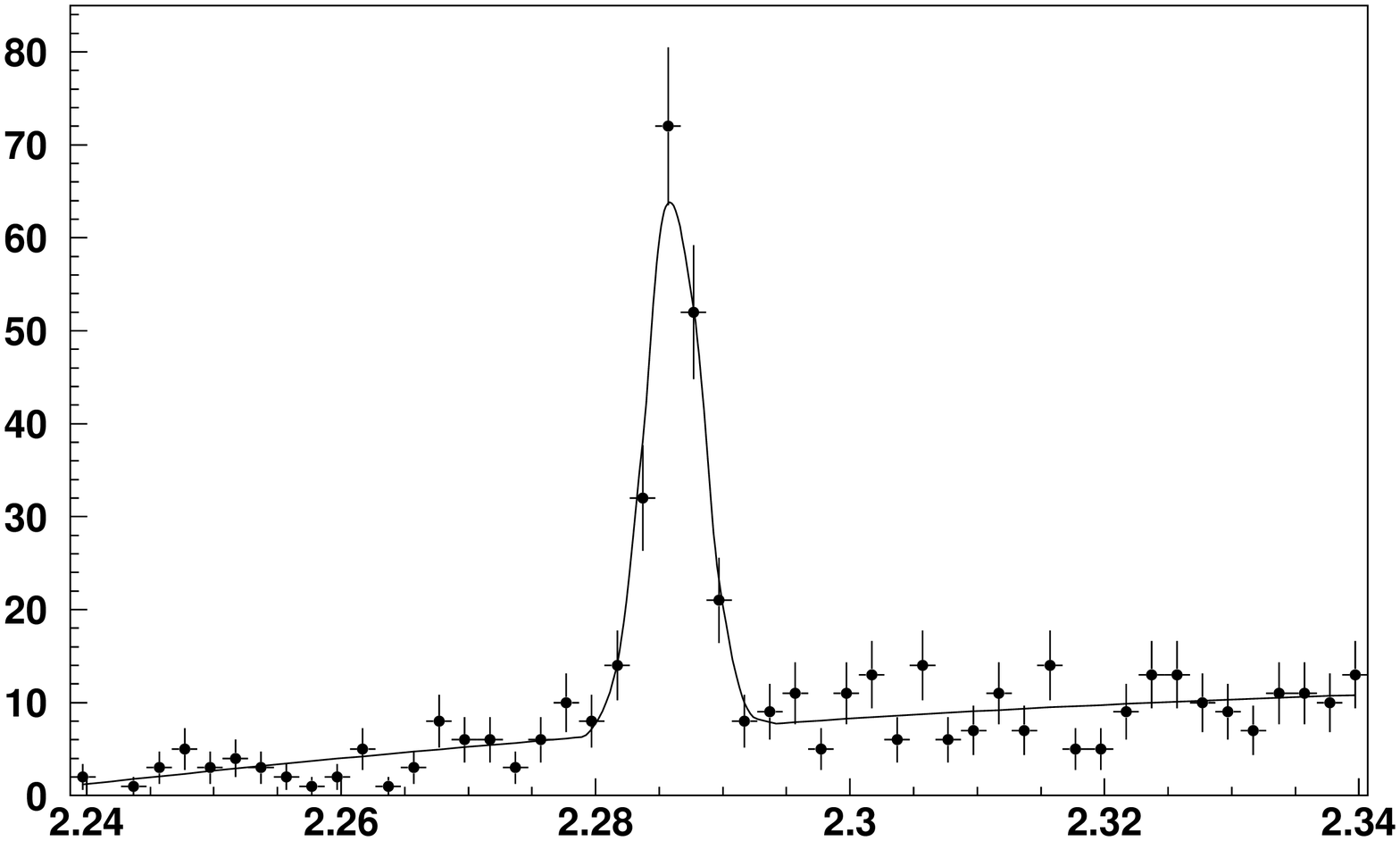}}
\end{picture}
\caption{$\Lambda_c^+ \ra \sig K^+ K^-$: invariant mass spectrum of the selected
	$\sig K^+ K^-$  combinations.}  
\label{fig_lamc_sigkk}
\end{figure}

\begin{figure}[p]
\centering
\begin{picture}(550,220)
\put(30,190){\large $\frac{N}{2.0~{\rm MeV}/c^2}$} 
\put(155,20){$M(\sig \pi^+ \pi^-)-M(\sig)+1.189,~{\rm GeV}/c^2$} 
\put(90,25){\includegraphics[width=0.7\textwidth]{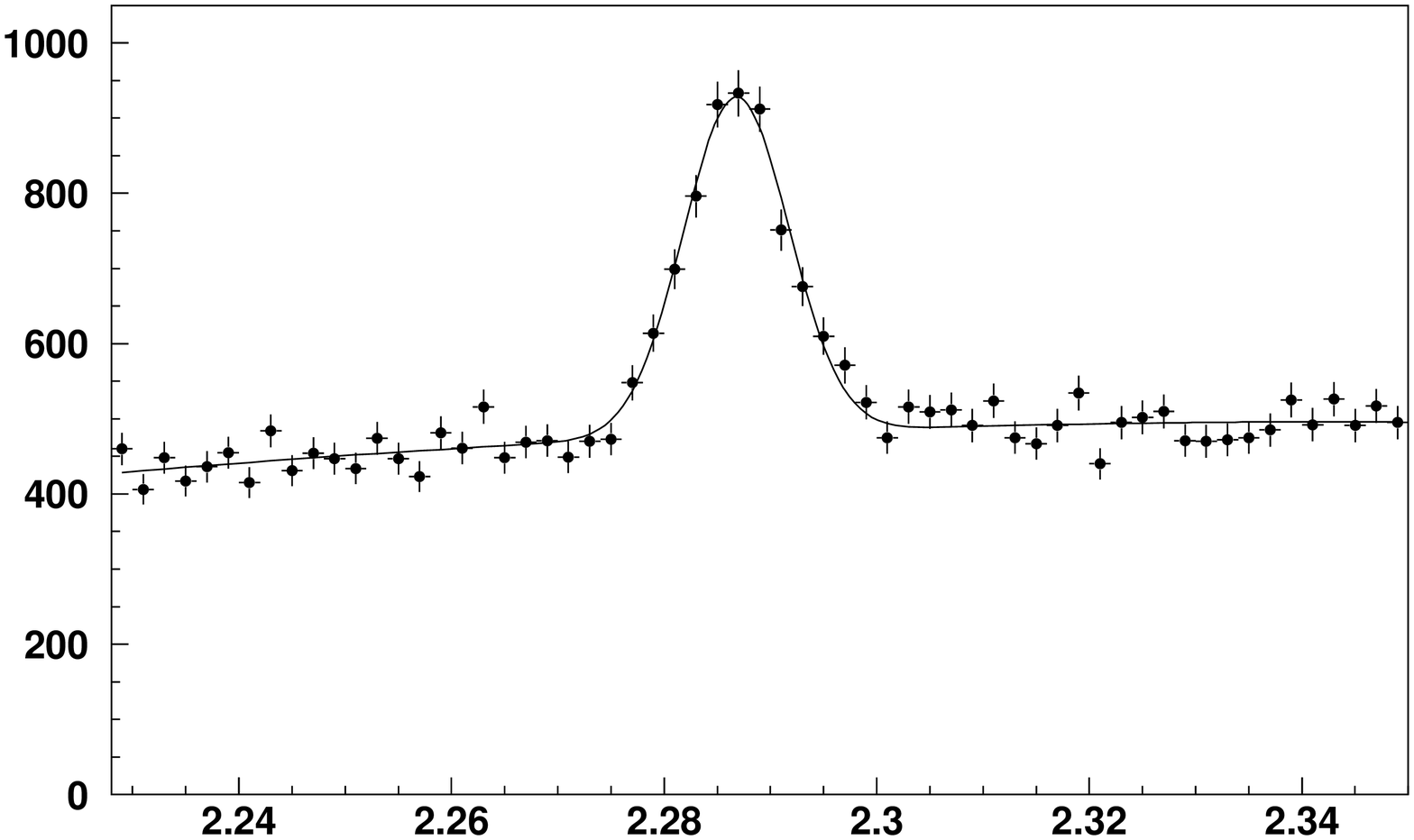}}
\end{picture}
\caption{The invariant mass spectrum for the normalization mode
	$\Lambda_c^+ \ra \sig \pi^+ \pi^-$, using equivalent cuts to the
	$\Lambda_c^+ \ra \sig K^+ K^-$ analysis.}  
\label{fig_lamc_sigpipi}
\end{figure}

The decays $\Lambda_c^+ \ra \sig K^+ K^-$ and $\Lambda_c^+ \ra \sig \phi$
proceed dominantly via W-exchange diagrams, and were observed by CLEO in
1993~\cite{CLEO_lamc_sigkk} with branching ratios
${\cal B}(\Lambda_c^+ \ra \sig K^+ K^-)/{\cal B}(\Lambda_c^+ \ra p K^- \pi^+) = 0.070 \pm 0.012 \pm 0.011$ and
${\cal B}(\Lambda_c^+ \ra \sig \phi)/{\cal B}(\Lambda_c^+ \ra p K^- \pi^+) = 0.069 \pm 0.023 \pm 0.016$.
Here we measure these decay channels with improved accuracy 
and provide the first evidence
for the $\Lambda_c^+ \ra \Xi(1690)^0 K^+$ decay.

Fig.~\ref{fig_lamc_sigkk} shows the invariant mass spectrum for
$\Lambda_c^+ \ra \sig K^+ K^-$ combinations selected according to
section~\ref{section-selection}, with the kaon cuts tightened to 
P($K/\pi) > 0.9$, and the impact parameter cut for the proton from
$\sig \ra p \pi^0$ relaxed to $d_{r\phi} > 200\, \mu\mathrm{m}$.
A clear peak is seen at the $\Lambda_c^+$ mass, over a low background.
We fit the distribution using a gaussian (with width fixed to 2.2~MeV$/c^2$
from the MC) plus a second order polynomial: the fit yields $161\pm 16$
 $\Lambda_c^+ \ra \sig K^+ K^-$ decays.
For normalization we reconstructed the $\Lambda_c^+ \ra \sig \pi^+ \pi^-$ decay
mode with equivalent cuts, shown in Fig.~\ref{fig_lamc_sigpipi}, and fitted
the distribution in the same manner: we find 
$2759\pm 138$ $\Lambda_c^+ \ra \sig \pi^+ \pi^-$ events.
The relative efficiency of the $\Lambda_c^+ \ra \sig K^+ K^-$ decay 
reconstruction with respect to the $\Lambda_c^+ \ra \sig \pi^+ \pi^-$ 
decay was calculated by MC simulation and was found to be $0.052/0.067 = 0.78$. 
We thus extract a branching ratio
\[
{\cal B}(\Lambda_c^+ \ra \sig K^+ K^-)/{\cal B}(\Lambda_c^+ \ra \sig \pi^+ \pi^-) = (7.5 \pm 0.8 \pm 1.5)\times 10^{-2},
\]
where the second (systematic) error is dominated by the uncertainty in the
relative $K/\pi$ identification efficiency.

In order to obtain the $\Lambda_c^+ \ra \sig \phi$ signal,
we take $\sig K^+ K^-$ from a $\pm 5$~MeV$/c^2$ window around the fitted
$\Lambda_c^+$ mass (2286~MeV$/c^2$), and plot the invariant mass of the 
$K^+ K^-$ combination, Fig.~\ref{fig_lamc_sigkk_mkk} (points with error bars);
the equivalent distribution is also shown for $\sig K^+ K^-$ in sidebands
centred 10~MeV$/c^2$ below and above the fitted $\Lambda_c^+$ mass
(shaded histogram).
The distributions are fitted with a Breit-Wigner function (describing the $\phi$ signal)
convoluted with a gaussian of a fixed width
(representing the detector mass resolution)
plus a 2nd order polynomial multiplied by a square root threshold factor.
The intrinsic width of the $\phi$ Breit-Wigner function was fixed to its
nominal value~\cite{PDG}, and the width of the gaussian resolution was
fixed to 1.0~MeV$/c^2$ based on the MC simulation.
The fit yields $106 \pm 12$ events for the $\phi$ signal in the $\Lambda_c^+$
region and $15 \pm 6$ in the $\Lambda_c^+$ sidebands. 
To extract the $\Lambda_c^+ \ra \sig \phi$ contribution we subtract the $\phi$
yield in the sidebands from the yield in the $\Lambda_c^+$ signal region,
correcting for the phase space factor obtained from the $\sig K^+ K^-$
background fitting function.
After making a further correction for the missing signal outside the
$\Lambda_c^+$ mass interval,
we obtain $93 \pm 14$ $\Lambda_c^+ \ra \sig \phi$ decays.

\begin{figure}[t!]
\centering
\begin{picture}(550,220)
\put(30,190){\large $\frac{N}{2.0~{\rm MeV}/c^2}$} 
\put(205,20){$M(K^+ K^-),~{\rm GeV}/c^2$} 
\put(90,25){\includegraphics[width=0.7\textwidth]{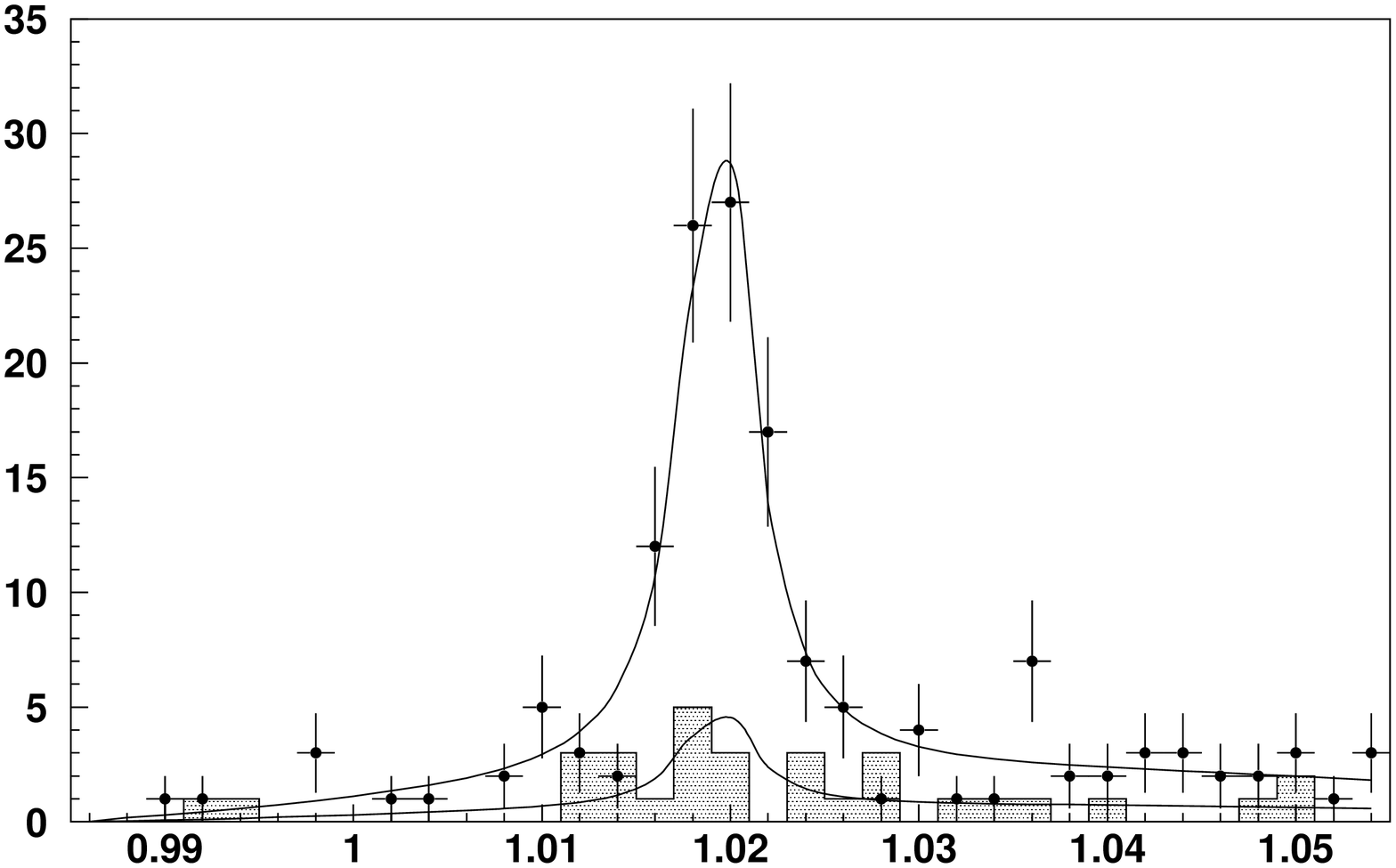}}
\end{picture}
\caption{Fitting for the $\Lambda_c^+ \ra \sig \phi$ component:
	the invariant mass spectra of $K^+ K^-$ combinations from
	the $\Lambda_c^+ \ra \sig K^+ K^-$ signal area (points with error bars)
	and $\Lambda_c^+$ sidebands (shaded histogram) are shown. 
	The selection requirements and fit are described in the text.}  
\label{fig_lamc_sigkk_mkk}
\end{figure}

The relative efficiency of the $\Lambda_c^+ \ra \sig \phi$ reconstruction
with respect to $\Lambda_c^+ \ra \sig \pi^+ \pi^-$ 
was calculated using the Monte Carlo and found to be $0.050/0.067 = 0.76$. 
Taking into account the $\phi$ branching fraction
${\cal B}(\phi \ra K^+ K^-) = 49.2\%$~\cite{PDG},
we calculate a branching ratio
\[
   {\cal B}(\Lambda_c^+ \ra \sig \phi)
  /{\cal B}(\Lambda_c^+ \ra \sig \pi^+ \pi^-)
  = (9.1 \pm 1.4 \pm 1.8) \times 10^{-2}.
\]
The systematic error of $1.8 \times 10^{-2}$ is an estimate based on
uncertainties in the relative $K/\pi$ identification efficiency, and fit
variations. In this case there is an additional source of systematic error due
to the difference in kinematics between the signal and normalization modes:
this has not yet been taken into account.

We also searched for the resonant structure in the $\sig K^-$ system in $\Lambda_c^+ \ra \sig K^+ K^-$
decays. Fig.~\ref{fig_lamc_sigkk_sigk} shows the $\sig K^-$ invariant mass spectra for $\sig K^+ K^-$
combinations in a $\pm 5$~MeV$/c^2$ interval around the fitted $\Lambda_c^+$
mass (data points):
we also required $| M(K^+ K^-) - m_\phi | > 10\, \mathrm{MeV}/c^2$ to suppress
$\phi \ra K^+ K^-$.
Also shown are $\sig K^-$ from $\sig K^+ K^-$ combinations
selected inside $\pm 5\, \mathrm{MeV}/c^2$ sideband intervals 10~MeV$/c^2$
below and above the fitted $\Lambda_c^+$ mass (shaded histogram).
The $\sig K^-$ mass distribution shows evidence for the $\Xi(1690)^0$
resonant state. In order to extract this resonant contribution the histograms
were fitted with a Breit-Wigner function (describing the $\Xi(1690)^0$ signal)
plus a 3rd order polynomial multiplied by a square root threshold factor.
The fit yields $52.5 \pm 15.0$ events for the $\Xi(1690)^0$ signal in the
$\Lambda_c^+$ region, with a fitted mass and width in good agreement with
previous measurements of the $\Xi(1690)^0$ parameters~\cite{PDG}.
To fit the sidebands, the function parameters were fixed to the central values
obtained from the signal fit, and the normalization was floated: a yield of
$7.2 \pm 2.8$ events was found.

The $\Lambda_c^+ \ra \Xi(1690)^0 K^+$ contribution was obtained by subtracting
the $\Xi(1690)^0$ yield in the sidebands from the yield in the $\Lambda_c^+$
signal region, correcting the sideband contribution using the 
phase space factor obtained from the $\sig K^+ K^-$ background fitting function.
After a further correction for the missing signal outside the
$\Lambda_c^+$ mass interval, we obtained
$45 \pm 15$ $\Lambda_c^+ \ra \Xi(1690)^0 K^+$ decays.
Assuming the relative efficiency for reconstruction of the
$\Lambda_c^+ \ra \Xi(1690)^0 K^+$ channel with respect to  
$\Lambda_c^+ \ra \sig \pi^+ \pi^-$ to be the same as for the inclusive
$\sig K^+ K^-$ mode, we find a combined branching ratio
\[
   \frac{{\cal B}(\Lambda_c^+ \ra \Xi(1690)^0 K^+)}
 	{{\cal B}(\Lambda_c^+  \ra \sig \pi^+ \pi^-)}
   \times {\cal B}(\Xi(1690)^0 \ra \sig K^-)
   = (2.1 \pm 0.7 \pm 0.4) \times 10^{-2};
\]
where we have neglected possible effects due to interference with 
$\Lambda_c^+ \ra \sig \phi$, \emph{etc.}.

\begin{figure}[t!]
\centering
\begin{picture}(550,220)
\put(30,190){\large $\frac{N}{4.0~{\rm MeV}/c^2}$} 
\put(155,20){$M(\sig K^-)-M(\sig)+1.189,~{\rm GeV}/c^2$} 
\put(90,25){\includegraphics[width=0.7\textwidth]{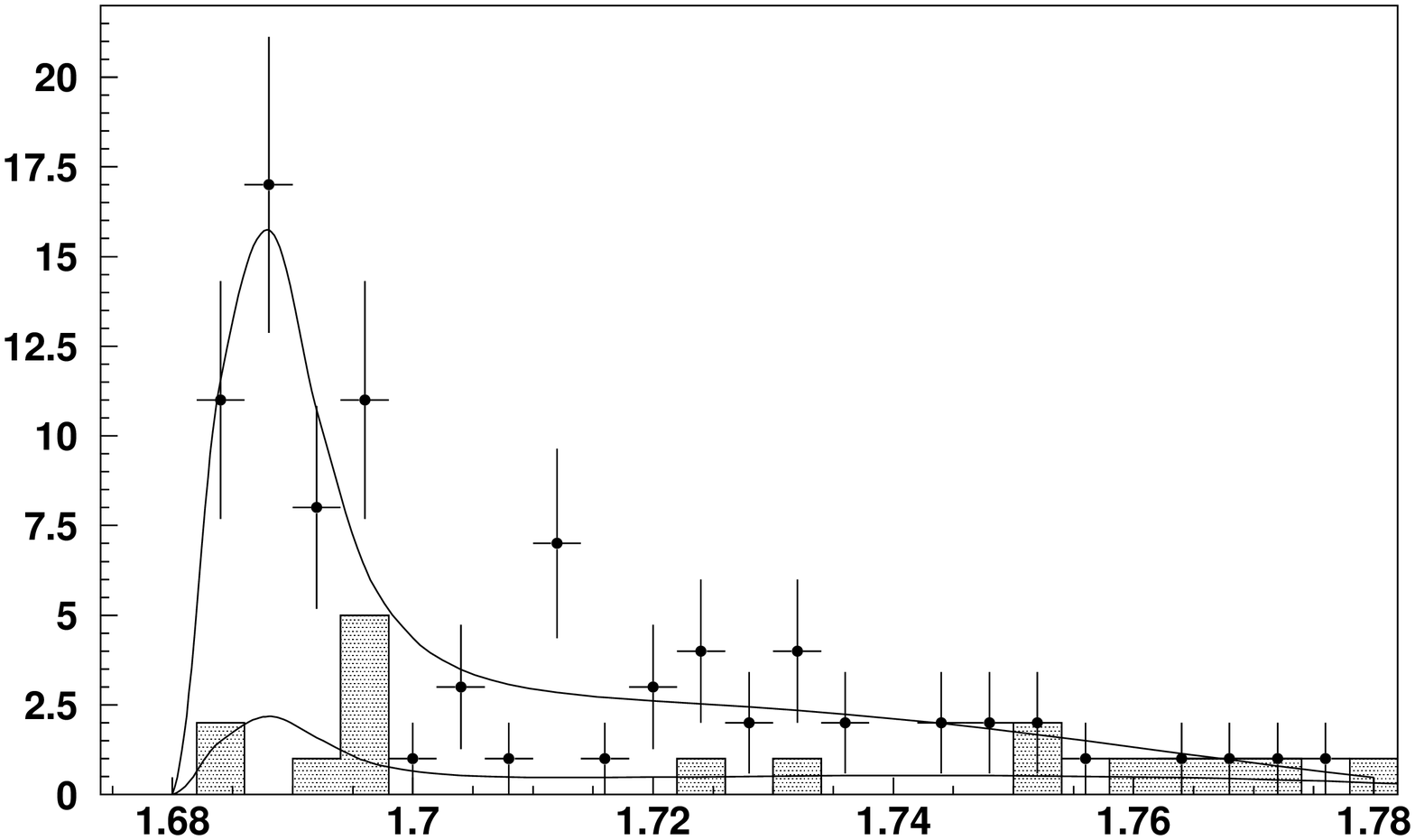}}
\end{picture}
\caption{Fitting for the $\Lambda_c^+ \ra \Xi(1690)^0 K^+$ component:
	the invariant mass spectrum of $\sig K^-$ combinations from the
	$\Lambda_c^+ \ra \sig K^+ K^-$ signal area (points with error bars) 
	and $\Lambda_c^+$ sidebands (shaded histogram) are shown, with
	the $\phi \ra K^+ K^-$ signal region excluded in both cases.
	The selection requirements and fit are described in the text.}  
\label{fig_lamc_sigkk_sigk}
\end{figure}

Finally, the non-resonant $\Lambda_c^+ \ra \sig K^+ K^-$ contribution is
estimated by making invariant mass cuts
$| M(K^+ K^-) - m_\phi |	> 10\, \mathrm{MeV}/c^2$ and
$| M(\sig K^-) - M_{\Xi(1690)}| > 20\, \mathrm{MeV}/c^2$ to suppress the
$\phi$ and $\Xi(1690)^0$ contributions (here, $M_{\Xi(1690)}$ is the fitted
$\Xi(1690)^0$ mass): the resulting $\sig K^+ K^-$ mass spectrum
is shown in Fig.~\ref{fig_lamc_sigkk_nonres}.
A fit with a gaussian (with width fixed to 2.2~MeV$/c^2$ from the MC) plus a
first order polynomial, shown on the figure, yields $23.8 \pm 7.1$ events.
Integrating the $\phi$ Breit-Wigner function over the allowed $M(K^+ K^-)$
region, we found that $14\%$ of the total $\Lambda_c^+ \ra \sig \phi$ signal
contributes to this sample: $12.3\pm 2.5$ events.
The contribution of the $\Xi(1690)^0$ mass tails is estimated to be
approximately 12$\%$ of the fitted $\Xi(1690)^0$ signal: $5.4 \pm 1.8$ events.
Subtracting these contributions, $6.1 \pm 7.7$ non-resonant events remain.
The phase space correction factor to account for the missing region around the
$\phi$ and $\Xi(1690)^0$ masses is found to be 1.63 by MC simulation of the
non-resonant $M(K^+ K^-)$  spectrum: applying this correction we obtain
$9.9 \pm 12.6$ $\Lambda_c^+ \ra \sig K^+ K^-$ non-resonant decays.
The relative efficiency for the reconstruction of the $\Lambda_c^+ \ra \sig K^+ K^-$ (non-resonant) 
channel with respect to the 
$\Lambda_c^+ \ra \sig \pi^+ \pi^-$ channel was taken to be the same as for the inclusive $\sig K^+ K^-$ mode.
Taking into account the systematic error on the $K/\pi$ identification
efficiency, we obtain an upper limit on the branching ratio
\[
   {\cal B}(\Lambda_c^+ \ra \sig K^+ K^-)_{non-res}
  /{\cal B}(\Lambda_c^+  \ra \sig \pi^+ \pi^-) 
  < 0.017
\]
at the 90\% confidence level, neglecting possible interference effects 
between $\Lambda_c^+ \ra \sig \phi$ and other resonant decays.

\begin{figure}[t!]
\centering
\begin{picture}(550,220)
\put(30,190){\large $\frac{N}{2.0~{\rm MeV}/c^2}$} 
\put(150,20){$M(\sig K^+ K^-)-M(\sig)+1.189,~{\rm GeV}/c^2$} 
\put(90,25){\includegraphics[width=0.7\textwidth]{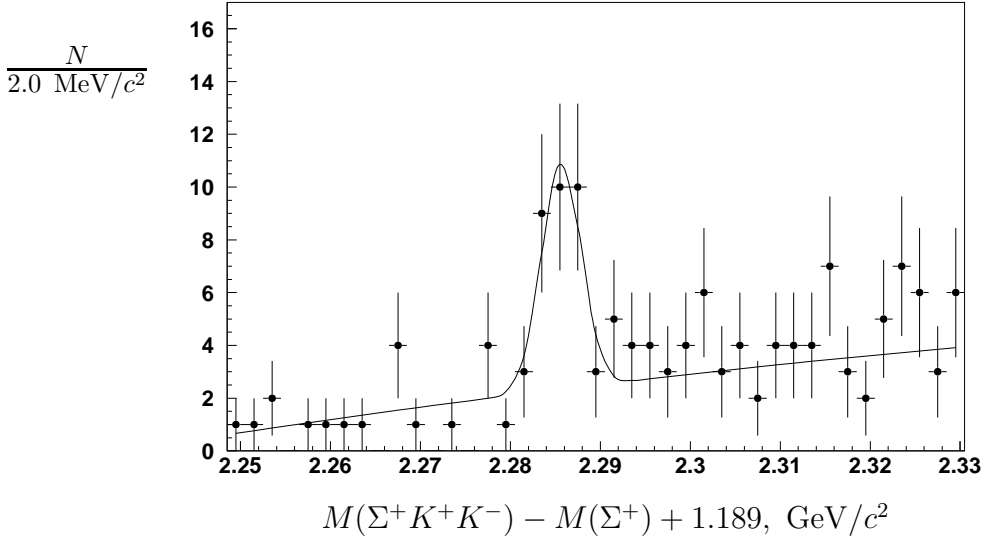}}
\end{picture}
\caption{The $\sig K^+ K^-$ invariant mass spectrum, after suppressing
	$\sig \phi$ and  $\Xi(1690)^0 K^+$ contributions:
	the selection requirements and fit are described in the text.}  
\label{fig_lamc_sigkk_nonres}
\end{figure}


\section{Measurement of the $\Lambda_{\lowercase {c}}^+ \ra {\lowercase {p}} K^+ K^-$ 
and $\Lambda_{\lowercase {c}}^+ \ra {\lowercase {p}} \phi$ decays}

\begin{figure}[p!]
\centering
\begin{picture}(550,220)
\put(30,190){\large $\frac{N}{2.0~{\rm MeV}/c^2}$} 
\put(210,20){$M(p K^+ K^-),~{\rm GeV}/c^2$} 
\put(90,25){\includegraphics[width=0.7\textwidth]{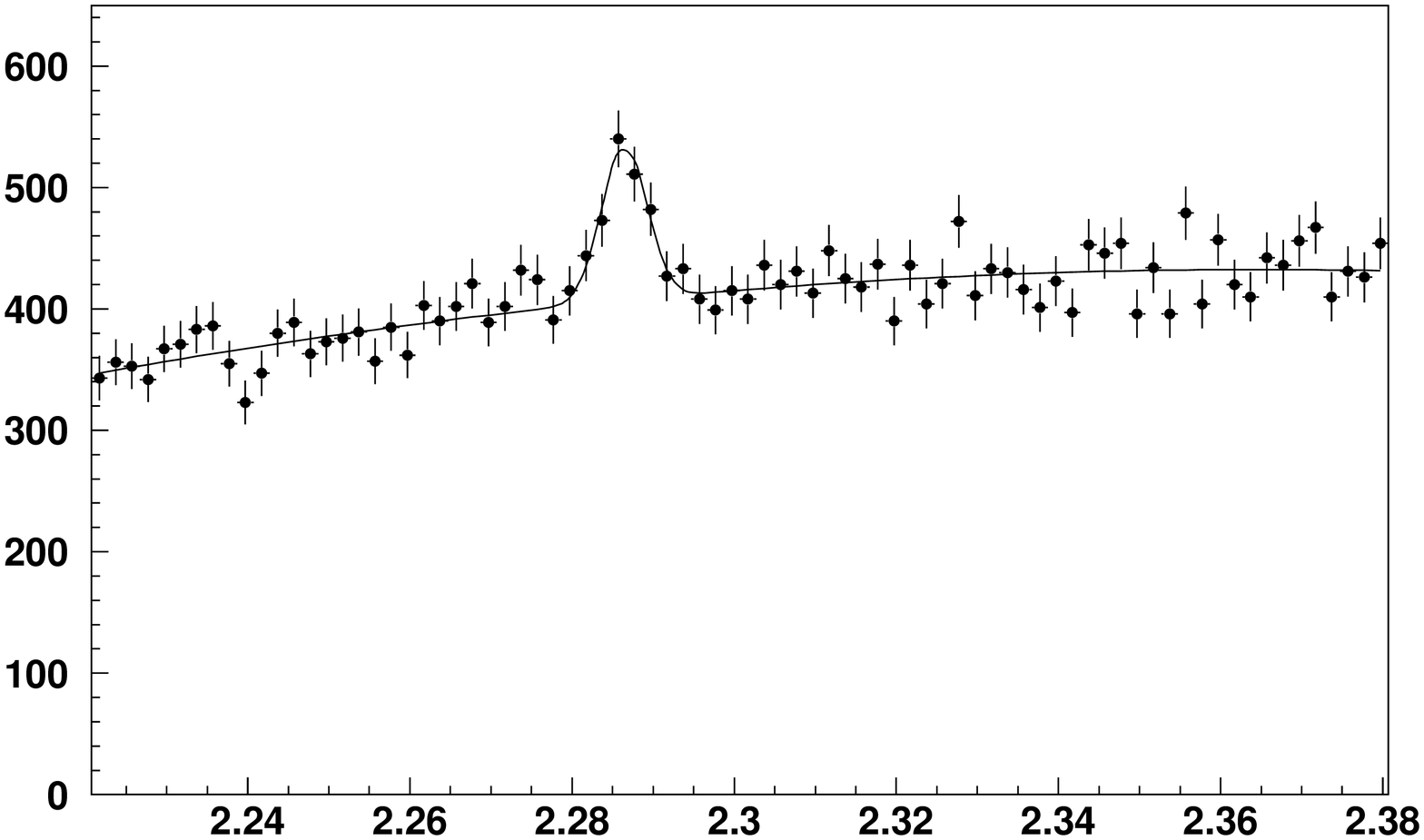}}
\end{picture}
\caption{$\Lambda_c^+ \ra p K^+ K^-$: invariant mass spectrum of the selected
	$p K^+ K^-$ combinations.} 
\label{fig_lamc_pkk}
\end{figure}

\begin{figure}[p!]
\centering
\begin{picture}(550,220)
\put(30,190){\large $\frac{N}{2.0~{\rm MeV}/c^2}$} 
\put(200,20){$M(p K^- \pi^+),~{\rm GeV}/c^2$} 
\put(90,25){\includegraphics[width=0.7\textwidth]{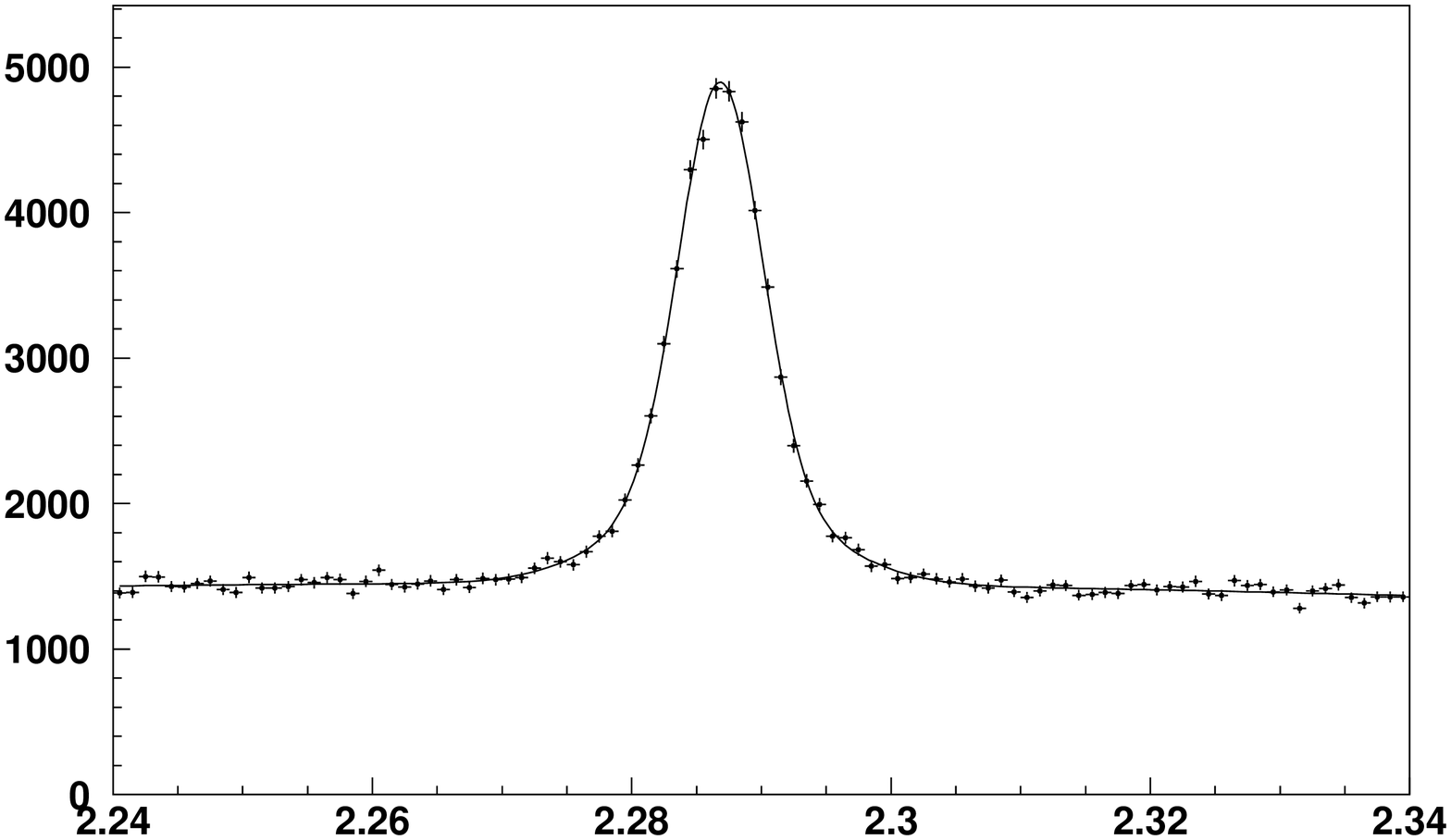}}
\end{picture}
\caption{The invariant mass spectrum for the normalization mode
	$\Lambda_c^+ \ra p K^- \pi^+$, using equivalent cuts to the
	$\Lambda_c^+ \ra p K^+ K^-$ analysis.}
\label{fig_lamc_pkpi}
\end{figure}

The first evidence for the $\Lambda_c^+ \ra p \phi$ decay was reported by NA32 in 1990, who claimed
a signal of $2.8 \pm 1.9$ events~\cite{NA32_lamc_pphi}.
The decay $\Lambda_c^+ \ra p K^+ K^-$ was observed for the first time by E687
in 1993, who also obtained an upper limit for the branching ratio of
$\Lambda_c^+ \ra p \phi$~\cite{E687_lamc_pkk}.
The most recent statistically significant resonant analysis was published by
CLEO in 1996, 
who found the following branching ratios:
${\cal B}(\Lambda_c^+ \ra p K^+ K^-)/{\cal B}(\Lambda_c^+ \ra p K^- \pi^+) = 0.039 \pm 0.009 \pm 0.007$
and ${\cal B}(\Lambda_c^+ \ra p \phi)/{\cal B}(\Lambda_c^+ \ra p K^- \pi^+) = 0.024 \pm 0.006 \pm 0.003$~\cite{CLEO_lamc_pkk}.

Reconstructing $\Lambda_c^+ \ra p K^+ K^-$ candidates according to 
section~\ref{section-selection}, with the kaon cuts tightened to 
P($K/\pi$)$> 0.9$, we see a clear peak at the $\Lambda_c^+$ mass, as shown
in Fig.~\ref{fig_lamc_pkk}. The tight P($K/\pi$) cut is necessary to suppress
the large combinatorial background.  We fit the distribution with a gaussian
(with width fixed to $2.8\,\mathrm{MeV}/c^2$ from the MC) plus a second order
polynomial, and find $446\pm 72$ $\Lambda_c^+ \ra p K^+ K^-$ events.
For normalization we reconstructed the $\Lambda_c^+ \ra p K^- \pi^+$ decay
mode with equivalent cuts (shown in Fig.~\ref{fig_lamc_pkpi}) and fitted the
distribution with a double gaussian for the large signal peak, and a second 
order polynomial, finding $33610\pm 1414$ events.
The relative efficiency of the $\Lambda_c^+ \ra p K^- K^+$ decay reconstruction
with respect to $\Lambda_c^+ \ra p K^+ \pi^-$ 
was found to be $0.21/0.24 = 0.88$ in the MC: using this value, we extract the
branching ratio 
\[
   {\cal B}(\Lambda_c^+ \ra p K^+ K^-)
  /{\cal B}(\Lambda_c^+ \ra p K^- \pi^+)
  = (1.50 \pm 0.25 \pm 0.15) \times 10^{-2}.
\]
The systematic error is dominated by the uncertainty in the relative $K/\pi$
identification efficiency.

In order to obtain the $\Lambda_c^+ \ra p \phi$ signal we take $p K^+ K^-$
from a $\pm 6\, \mathrm{MeV}/c^2$ ($\pm 2\sigma$) window
around the fitted $\Lambda_c^+$ mass (2286~MeV$/c^2$),
and plot the invariant mass of the $K^+K^-$ combination, 
Fig.~\ref{fig_lamc_pkk_mkk} (points with error bars); 
the equivalent distribution is also shown for $p K^+ K^-$ from
$6\, \mathrm{MeV}/c^2$ sideband intervals $10\, \mathrm{MeV}/c^2$ below and
above the fitted $\Lambda_c^+$ mass (shaded histogram).
The distributions are fitted with a Breit-Wigner function
(describing the $\phi$ signal)
convoluted with a gaussian of a fixed width
(representing the detector mass resolution)
plus a 3rd order polynomial multiplied by a square root threshold factor.
The width of the $\phi$ Breit-Wigner function was fixed to its nominal value~\cite{PDG},
and the width of the gaussian was fixed to $1.0\, \mathrm{MeV}/c^2$ based on MC simulation.
The fit yields $430 \pm 23$ events for the $\phi$ signal in the $\Lambda_c^+$
region and $232 \pm 17$ in the sidebands. 
To extract the $\Lambda_c^+ \ra p \phi$ contribution we subtract
the $\phi$ yield in the sidebands from the yields in the $\Lambda_c^+$ signal
region, correcting for the phase space factor obtained from the $p K^+ K^-$
background fitting function.
After making a further correction for the signal outside the $\Lambda_c^+$ mass
interval we obtain $205 \pm 30$ $\Lambda_c^+ \ra p \phi$ decays.

\begin{figure}[ht!]
\centering
\begin{picture}(550,220)
\put(30,190){\large $\frac{N}{2.0~{\rm MeV}/c^2}$} 
\put(205,20){$M(K^+ K^-),~{\rm GeV}/c^2$} 
\put(90,25){\includegraphics[width=0.7\textwidth]{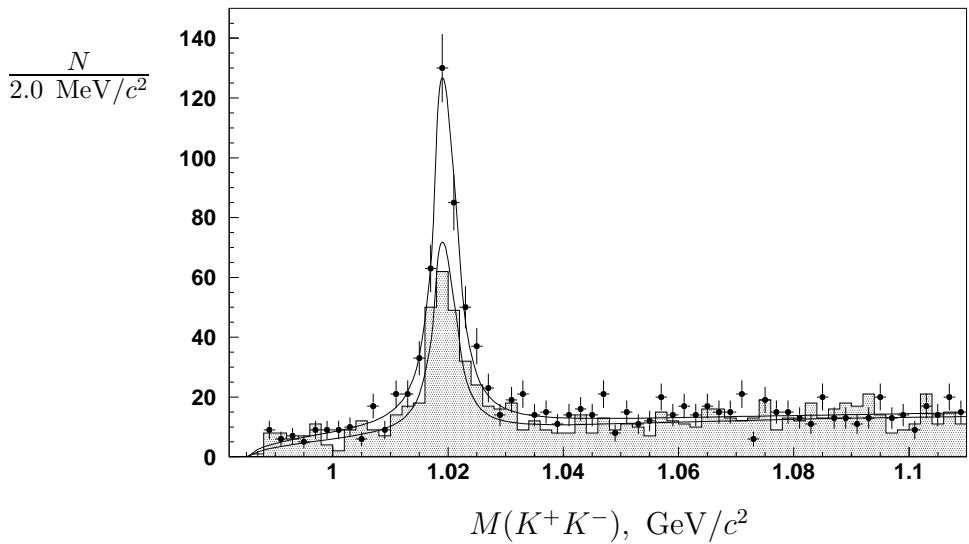}}
\end{picture}
\caption{Fitting for the $\Lambda_c^+ \ra p \phi$ component: 
	the invariant mass spectra of $K^+ K^-$ combinations from the
	$\Lambda_c^+ \ra p K^+ K^-$ signal area (points with error bars)
	and sidebands (shaded histogram).
	The selection requirements and fit are described in the text.}  
\label{fig_lamc_pkk_mkk}
\end{figure}

The reconstruction efficiency of the $\Lambda_c^+ \ra p \phi$ decay 
relative to $\Lambda_c^+ \ra p K^+ \pi^-$ was calculated using the Monte Carlo 
and found to be $0.20/0.24 = 0.83$. 
Taking into account the $\phi$ branching fraction
${\cal B}(\phi \ra K^+ K^-) = 49.2\%$~\cite{PDG},
we calculate a branching ratio 
\[
   {\cal B}(\Lambda_c^+ \ra p \phi)
  /{\cal B}(\Lambda_c^+ \ra p K^- \pi^+)
  = (1.50 \pm 0.23 \pm 0.15) \times 10^{-2}.
\]

\begin{figure}[ht!]
\centering
\begin{picture}(550,220)
\put(30,190){\large $\frac{N}{2.0~{\rm MeV}/c^2}$} 
\put(200,20){$M(p K^+ K^-),~{\rm GeV}/c^2$} 
\put(90,25){\includegraphics[width=0.7\textwidth]{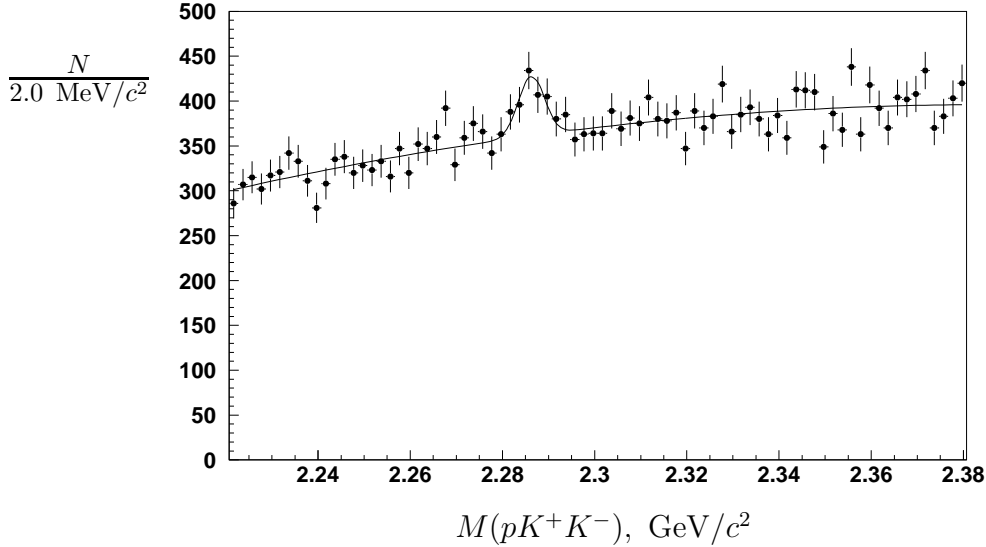}}
\end{picture}
\caption{The $\Lambda_c^+ \ra p K^+ K^-$ invariant mass spectrum,
	after suppressing the $p \phi$ contribution:
	the selection requirements and fit are described in the text.}  
\label{fig_lamc_pkk_nonphi}
\end{figure}

The non-$\phi$ $\Lambda_c^+ \ra p K^+ K^-$ signal is estimated by making an
invariant mass cut $| M(K^+ K^-) - m_\phi | > 10\, \mathrm{MeV}/c^2$ to suppress
the $\phi \ra K^+ K^-$ contribution: the resulting $p K^+ K^-$ mass spectrum is shown in Fig~\ref{fig_lamc_pkk_nonphi}.
A fit with a gaussian (with width fixed to $2.8\, \mathrm{MeV}/c^2$ from the 
MC) and a second order polynomial, shown on the figure,
yields $234 \pm 59$ events. 
Integrating the $\phi$ Breit-Wigner function over the allowed $M(K^+K^-)$
region, we found that $14\%$ of the total $\Lambda_c^+ \ra p \phi$ signal
($27.6\pm 6.2$ events) contributes: subtracting these, we have
$206 \pm 59$ events.
The phase space correction factor accounting for the missing region around the
$\phi$ mass is found to be 1.075 by MC simulation of the non-$\phi$
$M(K^+K^-)$  spectrum: applying this correction we obtain
$222 \pm 64$ $\Lambda_c^+ \ra p K^+ K^-$ non-$\phi$ decays .
Assuming the relative efficiency for reconstruction of
$\Lambda_c^+ \ra p K^+ K^-$ (non-$\phi$) decays with respect to
$\Lambda_c^+ \ra p K^- \pi^+$ to be the same as for inclusive $p K^+ K^-$,
we calculate a branching ratio 
\[
   {\cal B}(\Lambda_c^+ \ra p K^+ K^-)_{non-\phi}
  /{\cal B}(\Lambda_c^+  \ra p K^- \pi^+) 
  = (0.75 \pm 0.23 \pm 0.08) \times 10^{-2}.
\]
As with previous channels, the uncertainty in the relative $K/\pi$
identification efficiency dominates the systematic error.


\section{Conclusions}

In summary, we report the first observation of
the Cabibbo-suppressed decays $\Lambda_c^+\ra \lam K^+$ and
$\Lambda_c^+\ra \si K^+$,
and the first observation of $\Lambda_c^+\ra \Sigma^+ K^+\pi^-$ with
large statistics.
The decays $\Lambda_c^+ \ra p K^+ K^-$, 
$\Lambda_c^+\ra p \phi$ and $\Lambda_c^+\ra (p K^+ K^-)_{{\rm non-}\phi}$,  
and the W-exchange decays $\Lambda_c^+ \ra \sig K^+ K^-$ and
$\Lambda_c^+ \ra \sig \phi^0$ have been measured with the best accuracy to date.
We have also observed evidence for the decay $\Lambda_c^+ \ra \Xi(1690)^0 K^+$
and set an upper limit on the
non-resonant decay mode $\Lambda_c^+ \ra \sig K^+ K^-$.
The results for these decay modes are listed in Table 1:
all values are preliminary.

\begin{table}[bth]
\caption { Results obtained in this paper. }
\medskip
\begin{tabular}{lll}
{\bf Decay mode} & {\bf Ratio of branching fractions} & \\ 
\hline
$\Lambda_c^+\ra \lam K^+$ & ${\cal B}(\Lambda_c^+\ra \lam K^+)/{\cal B}(\Lambda_c^+\ra \lam \pi^+)$ & $=0.085\pm 0.012\pm 0.015$ \\
$\Lambda_c^+\ra \si K^+$ & ${\cal B}(\Lambda_c^+\ra \si K^+)/{\cal B}(\Lambda_c^+\ra \si \pi^+)$ & $=0.073\pm 0.018\pm 0.016$ \\
$\Lambda_c^+\ra \Sigma^+ K^+ \pi^-$  & ${\cal B}(\Lambda_c^+\ra \Sigma^+ K^+ \pi^-)/{\cal B}(\Lambda_c^+\ra \Sigma^+ \pi^+ \pi^-)$ & $=0.059 \pm 0.014 \pm 0.006$  \\
$\Lambda_c^+\ra \Sigma^+ K^+ K^-$ & ${\cal B}(\Lambda_c^+\ra \Sigma^+ K^+ K^-)/{\cal B}(\Lambda_c^+\ra \Sigma^+ \pi^+ \pi^-)$ & $=0.075 \pm 0.008 \pm 0.015$\\
$\Lambda_c^+\ra  \Sigma^+ \phi$ & ${\cal B}(\Lambda_c^+\ra \Sigma^+ \phi)/{\cal B}(\Lambda_c^+\ra \Sigma^+ \pi^+ \pi^-)$ & $=0.091 \pm 0.014 \pm 0.018$\\
$\Lambda_c^+\ra  \Xi(1690)^0 K^+$ & ${\cal B}(\Lambda_c^+\ra \Xi(1690)^0 K^+) \times$ & \\
 & $\times {\cal B}(\Xi(1690)^0 \ra  \sig K^-)/{\cal B}(\Lambda_c^+\ra \Sigma^+ \pi^+ \pi^-)$  & $= 0.021 \pm 0.007 \pm 0.004$\\
$\Lambda_c^+\ra (\Sigma^+ K^+ K^-)_{\rm non-res}$ & ${\cal B}(\Lambda_c^+\ra \Sigma^+ K^+ K^-)_{\rm non-res}/{\cal B}(\Lambda_c^+\ra \Sigma^+ \pi^+ \pi^-)$ & $<
0.017$ @ 90\% c.l.\\
$\Lambda_c^+\ra p K^+ K^-$ & ${\cal B}(\Lambda_c^+\ra p K^+ K^-)/{\cal B}(\Lambda_c^+\ra p K^- \pi^+)$ & $= (1.50 \pm 0.25 \pm 0.15) \times 10^{-2}$ \\
$\Lambda_c^+\ra p \phi$ & ${\cal B}(\Lambda_c^+\ra p \phi)/{\cal B}(\Lambda_c^+\ra p K^- \pi^+)$ & $= (1.50 \pm 0.23 \pm 0.15) \times 10^{-2}$ \\
$\Lambda_c^+\ra (p K^+ K^-)_{{\rm non-}\phi}$ & ${\cal B}(\Lambda_c^+\ra p K^+ K^-)_{{\rm non-}\phi}/{\cal B}(\Lambda_c^+\ra p K^- \pi^+)$ & $= (0.75 \pm 0.23 \pm 0.08) \times 10^{-2}$
\end{tabular}
\end{table}

\newpage 

\section{Acknowledgment}

We wish to thank the KEKB accelerator group for the excellent
operation of the KEKB accelerator. We acknowledge support from the
Ministry of Education, Culture, Sports, Science, and Technology of Japan
and the Japan Society for the Promotion of Science; the Australian
Research
Council and the Australian Department of Industry, Science and
Resources; the Department of Science and Technology of India; the BK21
program of the Ministry of Education of Korea and the CHEP SRC
program of the Korea Science and Engineering Foundation; the Polish
State Committee for Scientific Research under contract No.2P03B 17017;
the Ministry of Science and Technology of Russian Federation; the
National Science Council and the Ministry of Education of Taiwan; the
Japan-Taiwan Cooperative Program of the Interchange Association; and
the U.S. Department of Energy.


\end{document}